\documentclass[a4paper,11pt]{article}
\usepackage{jheppub} 
\usepackage{lineno}

\usepackage{macros}

\usepackage{amsthm}
\usepackage[T1]{fontenc}
\usepackage[utf8]{inputenc}
\DeclareUnicodeCharacter{2010}{-}

\usepackage{color}
\definecolor{dark-gray}{gray}{0.20}
\definecolor{gray}{gray}{0.30}
\definecolor{light-gray}{gray}{0.80}
\definecolor{dark-red}{rgb}{0.7,0,0}
\definecolor{dark-green}{rgb}{0.1,0.4,0}
\definecolor{dark-blue}{rgb}{0.3,0.3,0.7}
\definecolor{dark-blue2}{RGB}{0, 114,178}
\definecolor{light-blue}{rgb}{0.8,0.8,1}
\definecolor{swamp}{RGB}{240, 199, 197}
\definecolor{antiquefuchsia}{rgb}{0.57, 0.36, 0.51}

\usepackage{hyperref}
\usepackage{cleveref}
\hypersetup{
	colorlinks=true,
	linkcolor=dark-blue2,
	citecolor=dark-blue2,
	urlcolor=dark-blue2,
	linktoc=section
}

\usepackage{tocloft}

\title{Isolated Critical Points for Scherk-Schwarz Compactifications of M-theory}

\author{Zihni Kaan Baykara,}
\author{Shi Chen,}
\author{Cumrun Vafa}

\affiliation{Jefferson Physical Laboratory, Harvard University,
Cambridge, MA 02138, USA}

\emailAdd{zbaykara@g.harvard.edu, shichen@g.harvard.edu, vafa@g.harvard.edu}

\abstract{We consider Scherk-Schwarz compactifications of M-theory (toroidal compactifications with a non-trivial spin structure) in various dimensions and find isolated critical points of the potential on the moduli space. We demonstrate this by identifying the unbroken duality group and finding isolated points on the moduli spaces which are fixed by elements of the unbroken duality group. We work out concrete examples involving compactifications down to $d=8,6, 5$ and $4$ spacetime dimensions. We also conjecture a duality covariant anomaly cancellation condition for M-theory on $T^n$
 orbifolded by discrete $\mathbb Z_N$ symmetries acting as phases on the charge lattice.  This anomaly cancellation condition generalizes the level matching requirement for perturbative string orbifolds.
}
\begin{document}

\maketitle
\section{Introduction}
Most of the progress in constructing the string landscape has relied on the existence of supersymmetry.  However, we know that at least at low energies our universe is not supersymmetric, so those constructions do not directly relate to our universe.  In the absence of a full understanding of the non-supersymmetric landscape, one can still study the behavior of non-supersymmetric strings at weak coupling regions in the moduli space.  In such cases, however, one finds no stationary solutions of the potential. Indeed, it has been proposed that long-lived dS may not exist \cite{Obied:2018sgi}. It has also been argued as a Swampland principle \cite{Bedroya:2019snp,Bedroya:2019tba} (see also \cite{Rudelius_2019,Andriot:2022xjh}) that asymptotically at weak coupling 
\begin{align}
\Bigg|\frac{\nabla V}{V}\Bigg|^2\geq \frac{4}{d-2}
\end{align}
in reduced Planck units, which forbids critical points with $V>0$ in the weak coupling regime. Given that observations suggest $V>0$ while $|\nabla V| <V \sqrt{2}$ (see for example \cite{Agrawal:2018own,Bedroya:2025fwh}), we are led to study the behavior of $V$ in the interior of moduli space, where the coupling is no longer weak and we have no direct method to calculate $V$.  To find points where $|\nabla V|$ is small, it is natural to look for points where $\nabla V=0$ (presumably unstable) which by continuity would lead to small values of $|\nabla V|$ in that neighborhood.  When the moduli space contains points of enhanced gauge symmetry (typically involving a finite gauge symmetry) under which all moduli are charged, the potential is automatically critical at those points. This idea was originally suggested in \cite{Ginsparg:1986wr} and more recently it has been used in \cite{Chen:2025rkb,Baykara:2026gem,Baykara:2026vdc} to argue for the existence of critical points.  See also \cite{10.21468/SciPostPhys.15.6.224,ValeixoBento:2025yhz} for the use of other ingredients to get critical points and also \cite{Lust:2024aeg,Mohseni:2025tig} for recent work on the importance of global considerations to find non-supersymmetric vacua.

One of the simplest and oldest known ways to mildly break supersymmetry is the Scherk-Schwarz mechanism, where we compactify a higher dimensional theory on a circle with anti-periodic boundary conditions for fermions.   More generally we can compactify the theory on a torus with non-trivial spin structure to break supersymmetry.  It is natural to look for critical points in these cases. As expected, for large radii, where the computational techniques apply, one can show there are no critical points of $V$.  The main question is therefore how to argue for the existence of critical points in the interior of moduli space, where the radii are Planckian.

In this paper, we achieve this by determining the modified U-duality group of M-theory on a torus with non-trivial spin structures, which is a subgroup of the original U-duality group.  This is familiar from the study of NSR formulation of string theory where the non-trivial spin structure of the worldsheet torus breaks the $\SL(2,\ZZ)$ duality group of the worldsheet to a level 2 subgroup $\Gamma_0(2)$.  Our approach is to find the analog of the unbroken U-duality group when we have a non-trivial spin structure upon compactification of M-theory on $T^n$.  This leads us to find consistency conditions for orbifolds acting on M-theory charge lattice by phases generalizing the level matching conditions for string orbifolds.  Using this we show the methodology for finding isolated critical points for $V$.  For compactifications down to $d=8, 6, 5$ and $4$ dimensions, we explicitly locate such isolated critical points of $V$. For $7$ dimensions, we prove a no-go for such points.  We cannot determine the sign of $V$ at these critical points.

The organization of this paper is as follows:  
In section \ref{section2}, we first review U-duality groups for M-theory compactification on tori with trivial spin structure.  We then discuss the effect of making the spin structure non-trivial and explain how to characterize it in general U-duality frames (which we call \emph{generalized Scherk-Schwarz spin structure}).  In section \ref{section3}, we explain the moduli space of M-theory with generalized spin structure
and explain which subgroup of the U-duality group survives.  In section \ref{section4}, we show how one can use symmetries to find isolated critical points of the potential.  In section \ref{section5}, we provide examples of isolated critical points for compactifications to $8, 6, 5$ and $4$ dimensions.  We also show why this symmetry principle gives no critical points in $d=7$.  In section \ref{section6}, we end with concluding remarks. In appendix \ref{appendixA}, we formulate the quadratic conditions that correspond to consistency conditions for M-theory toroidal orbifolds acting as phases on the charge lattice. In appendix \ref{appendixB}, we study the symmetric points in the moduli space of positive definite lattices. We include supplementary Mathematica notebooks for the discussion and evaluation of computations.

\section{Scherk-Schwarz compactifications and duality groups}\label{section2}
\subsection{U-duality group}
One of the most well-studied quantum gravity theories is the maximal supergravity constructed by the toroidal compactification of M-theory. In this section, we review some basic properties of the moduli space, duality groups and the charge lattices of these theories.

We consider M-theory compactified on $T^n$ and denote the spacetime dimension as $d=11-n$. Maximal supersymmetry imposes strong constraints on its marked moduli space such that it must be a homogeneous space of the form \cite{Obers:1998fb}:\footnote{By the marked moduli space, we mean the moduli space prior to imposing identifications under the duality group, marked by the vevs of the scalar fields.}
\begin{equation}\label{coarse moduli space}
    \widehat{\mathcal{M}}_{n}=\E_{n(n)}(\RR)/\mathrm{H},
\end{equation}
with $\E_{n(n)}(\RR)$ the split real form of the exceptional Lie group and $\mathrm{H}$ its maximal compact subgroup, which can also be viewed as the R-symmetry group of the maximal supersymmetry.\footnote{There is an extra $\RR^+$ factor for $n=1,2$.} The explicit form of the space can be found in Table \ref{table of CMS}.
\begin{table}
    \centering
    \begin{tabular}{|c|c|}
    \hline
         $n$& Marked Moduli Spaces for M-theory on $T^{n}$ \\
         \hline
         1&$\RR^+$\\
         \hline
         2&$\RR^{+}\times\frac{\SL(2,\RR)}{\SO(2)}$\\
         \hline
         3&$\frac{\SL(3,\RR)}{\SO(3)}\times\frac{ \SL(2,\RR)}{\SO(2)}$\\
         \hline
         4&$\frac{\SL(5,\RR)}{\SO(5)}$\\
         \hline
         5&$\frac{\SO(5,5,\RR)}{\SO(5)\times\SO(5)}$\\
         \hline
         6&$\frac{\E_{6(6)}(\RR)}{\USp(8)}$\\
         \hline
         7&$\frac{\E_{7(7)}(\RR)}{\SU(8)}$\\
         \hline
         8&$\frac{\E_{8(8)}(\RR)}{\SO(16)}$\\
         \hline
    \end{tabular}
    \caption{The marked moduli spaces for M-theory compactified on $T^n$.}
    \label{table of CMS}
\end{table}

The low energy EFT for these theories has the symmetry group $\E_{n(n)}(\RR)$, but charge quantization in the non-perturbative string/M-theory UV completion breaks this continuous symmetry to an arithmetic subgroup, called the U-duality group:
\begin{equation}\label{U-duality group}
    \E_{n(n)}(\RR)\mapsto\E_{n(n)}(\ZZ).
\end{equation}
The best way to explain why this is the correct duality group is to note that $\E_{n(n)}(\ZZ)$ is generated by two subgroups
\begin{equation}
    \E_{n(n)}(\ZZ)=  \SL(n,\ZZ) \bowtie\Spin(n-1,n-1,\ZZ).
\end{equation}
The first subgroup $\SL(n,\ZZ)$ is the large diffeomorphism group of the torus $T^n$. If we decompose $T^n$ by $S^1\times T^{n-1}$, the theory becomes type IIA theory compactified on $T^{n-1}$, and the $\Spin(n-1,n-1,\ZZ)$ is the T-duality group of the perturbative type IIA superstring. Thus, the combination of the M-theory picture and the string theory picture implies that $\E_{n(n)}(\ZZ)$ is the U-duality group.

With the knowledge of the duality group, the moduli space is then the orbifold of the marked moduli space quotiented by the duality group:
\begin{equation}\label{moduli space}
    \mathcal{M}_n =\E_{n(n)}(\ZZ)\backslash\E_{n(n)}(\RR)/\mathrm{H}.
\end{equation}

The $0$-form/particle charges for the $n\leq 7$ theories are generated by KK momenta, M2-brane wrapping 2-cycles, M5-brane wrapping 5-cycles and KK monopoles. They form an integral representation of the U-duality group which can be summarized as in Table \ref{charge lattices}.
\begin{table}
    \centering
    \begin{tabular}{|c|c|c|c|}
    \hline
         $n$& U-duality group& Dimension &  Representation \\
         \hline
         1&1&$\mathbf{1}$ & \text{trivial}\\
         \hline
         2&$\SL(2,\ZZ)$&$\mathbf{2}\oplus\mathbf{1}$ & fundamental $\oplus$ trivial\\
         \hline
         3&$\SL(3,\ZZ)\times\SL(2,\ZZ)$&$(\mathbf{3},\mathbf{2})$& \text{bi-fundamental} \\
         \hline
         4&$\SL(5,\ZZ)$&$\mathbf{10}$ & anti-symmetric two tensor\\
         \hline
         5&$\Spin(5,5,\ZZ)$&$\mathbf{16}$ & spinor \\
         \hline
         6&$\E_{6(6)}(\ZZ)$&$\mathbf{27}$ & fundamental\\
         \hline
         7&$\E_{7(7)}(\mathbb{Z})$&$\mathbf{56}$ &  fundamental\\
         \hline
         8&$\E_{8(8)}(\mathbb{Z})$&$\mathbf{248}$ &adjoint\\
         \hline
    \end{tabular}
    \caption{The charge lattices as representations with respect to the U-duality groups for M-theory on $T^n$.}
    \label{charge lattices}
\end{table}

\subsection{Scherk-Schwarz compactification}
\subsubsection{Classical Scherk-Schwarz compactification}
To get a phenomenologically interesting model, we must break the supersymmetry by some mechanism. The simplest way to break supersymmetry for the toroidal compactification is the Scherk-Schwarz compactification \cite{Scherk:1979zr}. Namely, we put a non-trivial spin structure on the torus, so the spinors have non-trivial boundary conditions. Since the large diffeomorphism group acts transitively on the non-trivial spin structures of a torus, without loss of generality, we can choose
\begin{equation}\label{anti-periodic torus}
    T^{n}_A=S_A^1\times T^{n-1},
\end{equation}
where $S_A^1$ is a circle with an anti-periodic boundary condition for fermions and the rest of the directions are periodic. Equivalently, for $n\geq 2$, we can view this as type IIA string theory compactified on
\begin{equation}\label{anti-periodic torus II}
    T^{n-1}_A=S_A^1\times T^{n-2}.
\end{equation}
Clearly, the twisted spin structure kills the fermionic zero modes, so the resulting theories do not have any massless fermionic degrees of freedom. In particular, it would be non-supersymmetric. Also, the string theory calculation \cite{Rohm:1983aq} shows that this theory develops tachyonic winding modes for small radius of the anti-periodic circle.

It is easy to see that, at weak coupling, the scalar potential $V$ will not have any critical point.  So, to find critical points, we need to study $V$ at strong coupling regions. However, the lack of supersymmetry makes it extremely difficult to get information in these regions. In this paper, we want to use the duality group to get some non-trivial constraint on the theory. So, the first question to ask is how the U-duality groups interact with the existence of a non-trivial spin structure, which we now turn to.

\subsubsection{U-duality action on the spin structure}
We aim to find the proper U-duality group relevant for Scherk-Schwarz compactifications. Since the U-duality group is generated by a combination of the large diffeomorphism in the M-theory picture and the T-duality in the string theory picture, we can separate this problem into two parts: how do the large diffeomorphism group $\SL(n,\ZZ)$ and the T-duality group $\Spin(n-1,n-1,\ZZ)$ act on the spin structure of the torus?

First, we treat the torus with spin structure as the 
orbifold by the free $\ZZ_2$ action generated by the geometric half-shift $u$, namely, a translation on the torus such that $2u\equiv 0$ and acting on the torus coordinate $z$ as
\begin{align}
    z \mapsto z+u,
\end{align}
dressed with $(-1)^F$ acting as a minus sign on the spinor bundle
\begin{equation}\label{geo-orbifold}
    T^n_A=\frac{\tilde T^n}{(-1)^F\circ u},
\end{equation}
where $\tilde T^n$ denotes the covering torus, which has twice the radius as $T^n$ on the shifted direction.

Then, for a large diffeomorphism of the covering torus
\begin{equation}
    f:\tilde{T}^n\to \tilde{T}^{n},
\end{equation}
represented by $A\in\SL(n,\ZZ)$. It acts on the shift vector $u$ by
\begin{equation}
    u\mapsto Au,
\end{equation}
as the fundamental representation of $\SL(n,\ZZ)$. Thus, we get the M-theory picture:
\\\\
\emph{In the M-theory picture, fixing the covering torus $\tilde{T}^n$, the quotient tori with non-trivial spin structure are parametrized by non-zero half-shift vectors $u\in\ZZ_2^n$, and the large-diffeomorphism group of the covering torus $\tilde{T}^n$ acts on them as the fundamental representation.}
\\\\
In addition to the large diffeomorphisms descending from the covering torus $\tilde{T}^n$, there are also large diffeomorphisms of the quotient torus $T_A^n$ which cannot be lifted to a diffeomorphism of the covering. However, our ultimate goal, as we will explain in the next section, is to study the subgroup of the large diffeomorphism group \emph{fixing the spin structure}. It can be proven that there is a one-to-one correspondence between the large diffeomorphisms of $T^n_A$ fixing its spin structure and the large diffeomorphisms of $\tilde{T}^n$ fixing the shift vector $u$, so it is enough to direct our attention to the covering torus.

Now, we consider the type IIA picture and the T-duality group. In order to get a type IIA description, we first choose an ordinary circle in $T^n$ to do the reduction. The reduction will then correspond to type IIA string theory on $T^{n-1}$. The T-duality group $\Spin(n-1,n-1,\ZZ)$ is generated by the large diffeomorphisms $\SL(n-1,\ZZ)$ and T-dualities on each circle. We then need to see how this perturbative duality group interacts with the non-trivial spin structure.

We again treat the torus with spin structure as a perturbative string orbifold by a half-shift $u$ dressed with $(-1)^F$ as
\begin{equation}\label{orbifold}
    T^{n-1}_A=\frac{\tilde T^{n-1}}{(-1)^F\circ T_u},
\end{equation}
where $T_u$ is the action of shift $u$ on the worldsheet Hilbert space, and $\tilde T^{n-1}$ denotes the covering torus.

If we parametrize the states by the momentum and winding $| n, w\rangle$ in type IIA compactified on $\tilde{T}^{n-1}$, the operator $T_u$ acts as
\begin{equation}\label{T_d action}
    T_u| n, w\rangle=(-1)^{u\cdot n}| n, w\rangle
\end{equation}

Taking the vector $ N=( n, w)$ and representing $v=(0,u)$, we can rewrite \eqref{T_d action} more concisely
\begin{equation}\label{T_d action 2}
    T_{  v}|  N\rangle=(-1)^{  v\cdot    N}|  N\rangle,
\end{equation}
where the dot product uses the hyperbolic metric on the Narain lattice. More generally, $v$ can also have components dual to winding charges.\footnote{These components correspond to $\mathbb Z_2\subset U(1)$ gauge transformations associated with the $B_{\mu i}$ field.}

An arbitrary shift of this type need not satisfy level matching and may therefore suffer from a global anomaly. However, if the shift is dual to a geometric shift through a sequence of T-dualities and diffeomorphisms, then level matching is satisfied and the orbifold is unobstructed.

With this notation, let $g$ be an element of the T-duality group $\Spin(n-1,n-1,\ZZ)$, then it acts on the vector $  N$ as the vector representation
\begin{equation}\label{T-duality}
    g|  N\rangle = |g  N\rangle.
\end{equation}
Therefore, the T-duality acts on the  shift by
\begin{equation}
    g T_{v}g^{-1}|  N\rangle=g(-1)^{  v\cdot g^{-1}  N}| g^{-1}  N\rangle=(-1)^{(g  v)\cdot  N}|  N\rangle=T_{g   v}|  N\rangle
\end{equation}
Thus, the shift $v$ transforms in the vector representation of $\Spin(n-1,n-1,\ZZ)$ under the $\mathbb{Z}_2$ reduction.\footnote{In general, there can also be phases in the action $g$ on the Hilbert space, but they will cancel out for our purposes.} Since the $(-1)^F$ commutes with the $\Spin(n-1,n-1,\ZZ)$ transformation, we get the string theory picture:
\\\\
\textsl{In the type IIA picture, fixing the covering torus $\tilde{T}^{n-1}$, the spin structure is parametrized by a non-zero vector $v\in \ZZ_2^{2(n-1)}$, which acts on the momentum-winding states as a phase shift, and the T-duality group of the covering torus $\tilde{T}^{n-1}$ acts on it as the vector representation}
\\\\
We have seen that spin structures transform naturally under both large diffeomorphisms and T-duality of \emph{the covering torus}. Since these generate the full U-duality group, the latter should also act on the space of spin structures. However, in the type IIA frame, a spin structure need not be purely geometric: it may also involve a winding shift symmetry. This motivates a more general notion of spin structure that is compatible with the U-duality group.
\subsubsection{Generalized Scherk-Schwarz spin structure}
The conclusion of the previous section is the following: fixing a covering torus $\tilde{T}^n$, the torus with non-trivial spin structure generated from it is parametrized by
\begin{itemize} 
\item A geometric $\ZZ_2$ shift of $\tilde{T}^n$ from the M-theory picture; 
\item Under any reduction to a type II frame, it is the geometric-winding  $\ZZ_2$ shift in the perturbative string states. \end{itemize}

The natural frame independent generalization is a $\ZZ_2$ subgroup of the $\U(1)^{r_n}$ gauge group of M-theory compactification, where $r_n$ is the rank of the gauge group after compactifying on $\tilde {T}^n$.

We first set a basis convention for the charge lattice $\Lambda_n$. The first entries correspond to KK momenta $p_i$, followed by M2 winding charges $w^{ij}$, and, if present, M5 winding charges $s^{ijklm}$ and KK monopole charges $t^i$ on the cycles generated by the basic circles of $\tilde{T}^n$ as
\begin{align}
    q =(p_i, w^{ij}, s^{ijkl m} ,t^{i})\in \Lambda_n.
\end{align}

A $\ZZ_{N}$ discrete gauge transformation, acting as phases on the charge lattice, is then parametrized by a $\ZZ_N$ \emph{phase shift vector} $v$ in the dual lattice $\Lambda_{n,N}^{*}:=\Hom(\Lambda_n,\ZZ_{N})$, which we call the \emph{phase shift lattice}. We then expand $v\in \Lambda^*_{n,N}$ by the dual basis:
\begin{align}
    v= (u^i, \omega_{ij}, \sigma_{ijklm},\tau_i) \in \Lambda_{n,N}^*.
\end{align}
Then given a $\ZZ_2$ phase shift vector $v$, we can describe the torus with a spin structure as an orbifold background:
\begin{equation}\label{General orbifold}
    T^n_A=\frac{\tilde{T}^{n}}{(-1)^{F}\circ T_v}.
\end{equation}
where $T_v$ is the $\ZZ_2$ discrete gauge transformation operator corresponding to the phase shift vector $v$. The geometric half-shift and the winding shift will then become a special case for this more general description with
\begin{align}\label{momentum-winding}
    v= (u^i,\omega_{i1},0,0),
\end{align}
for $i=2,\cdots,n$, where $\omega_{i1}$ corresponds to the shift on the charge of M2 wrapping the $1-i$ cycle, which becomes F1 winding charge after the M-theory reduction on circle $1$.
We conclude
\\\\
\emph{The spin structure is parametrized by a non-zero phase shift vector $v\in \Lambda_{n,2}^*$, where $\Lambda_{n,2}^*\simeq \ZZ_2^{r_n}$ is the lattice of $\ZZ_2$ phase shift vectors inside $\U(1)^{r_n}$ Cartan subgroup of the gauge symmetries of M-theory on $\tilde{T}^n$.}
\\\\
More generally, for a $\mathbb Z_N$ phase shift vector $v$, it is not the shift vector itself, but the $\mathbb Z_N$ subgroup it generates in $\Lambda_{n, N}^*\simeq\mathbb Z_N^{r_n}$ that determines the orbifold group. Therefore, the correct parameter space of $\ZZ_{N}$ phase shift orbifold groups is
\begin{align}
    [v]\in \mathbb P(\Lambda_{n,N}^*)=\PP(\ZZ_N^{r_n}),
\end{align}
where the projective space $\PP(\ZZ_{N}^r)$ is the space of $\ZZ_{N}$ subgroups in $\ZZ_{N}^{r}$. Therefore, in the cases of general $N$, we have the following statement:
\\\\
\emph{A general $\ZZ_N$ phase shift group is parametrized by an element $[v]$ in $\PP( \Lambda_{n,N}^*)$.}
\\\\
For $N=2$, the $\ZZ_2$ subgroup generated by $v$ can be simply labeled by the non-zero $v$ itself; thus, the two descriptions coincide.

However, as mentioned before, not all choices of $ v$ can lead to an anomaly-free orbifold. Consider the perturbative string orbifold
\begin{equation}
    T^{n-1}_A=\frac{\tilde T^{n-1}}{(-1)^{F}\circ T_{ v}},
\end{equation}
with $v$ of the form in \eqref{momentum-winding}, which acts as a momentum-winding shift. The level matching condition needed for modular invariance \cite{Vafa:1986wx} is
\begin{equation}\label{level-matching}
    \frac{1}{2} v \cdot v =0 \mod 2.
\end{equation}
The integral quadratic form $ \frac{1}{2} v \cdot v $ is the defining quadratic form of $\Spin(n-1,n-1, \ZZ_2)$.\footnote{For characteristic $2$ fields, the notions of the symmetric bilinear form and the quadratic form are not equivalent. In particular, $\SO(n-1,n-1, \ZZ_2)$ is the group of matrices preserving the quadratic form $q(v)=\frac 12 v\cdot v$, not the bilinear form $v\cdot w$.} But this means that $ v\in  \ZZ_2^{2(n-1)}$ is a $ \ZZ_2$ null vector. Since the group $\Spin(n-1,n-1, \ZZ_2)$ acts transitively on the null vectors, we get the following conclusion:
\\\\
\emph{The $\ZZ_2$ phase shift $v$ in perturbative string theory has a consistent orbifold if and only if it is in the T-duality orbit of a geometric shift.}
\\\\
Now, considering a general phase shift vector $v$, we know that it gives us a consistent string theory orbifold background if, by using a U-duality transformation, it can be mapped to a geometric shift.  Thus, we give the following definition:
\\\\
\emph{A \textbf{generalized Scherk-Schwarz spin structure} is a choice of a $\ZZ_2$ phase shift vector $v\in \PP( \Lambda_{n,2}^*)$ that is in the U-duality orbit of a geometric shift.}
\\\\
An important clarification is that the vector $v$ does not parametrize non-trivial spin structures on a \emph{fixed} torus background $T^n_A$. Instead, it parametrizes all tori with non-trivial spin structure $T_A^n$ obtained by orbifolding the \emph{fixed} covering torus $\tilde{T}^n$. Note that these orbifolds are not isometric to each other in general. Therefore, what we have done is study how the large diffeomorphism group, T-duality group, hence the U-duality group of the covering torus acts on the generalized spin structure $v$.

For $n\leq 4$, the orbit structure of the lattice of phase shifts under the U-duality group is relatively simple. In appendix \ref{appendixA}, we show that $\ZZ_2$ shifts that are not in the orbit of a geometric shift can be represented by an anomalous $\ZZ_2$ geometric-winding shift in the perturbative string theory; therefore, it does not produce a consistent orbifold. Hence, in this paper, we restrict our discussion only to the U-duality orbit of geometric spin structures.

The natural question to ask is then the following: given an element $ v\in  \Lambda_{n,2}^{*}$, how do we know if it is in the orbit of the geometric shift? Using the fact that the U-duality group is a split semi-simple algebraic group and the charge lattice is a \emph{minuscule representation}, we can apply the techniques of algebraic geometry (see appendix \ref{appendixA}) to prove that the orbit of the geometric spin structure is a projective algebraic variety inside $\PP(  \Lambda_{n,2}^{*})$ defined by a set of quadratic equations listed in Table \ref{quadratic-eqs}. For the explicit form of the equations, see appendix \ref{Characterizing Equations}. In other words, there exists a set of quadratic equations $f_i(x_1,\cdots,x_{r_n})$ such that:
\begin{equation}
    \text{$ v\in\PP(  \Lambda_{n,2}^{*})$ is in the orbit of the geometric shift}\Leftrightarrow f_i( v)=0.
\end{equation}
We then denote the U-duality orbit of geometric shifts by $\mathcal{S}_n$, which is the set of generalized Scherk-Schwarz spin structures of M-theory on $\tilde{T}^n$.
\begin{table}[]
    \centering
    \begin{tabular}{|c|c|}
    \hline
         $n$ & Equation \\
         \hline
         3 & $\det v = 0$ \\
         4 & $\frac{1}{2}v \wedge v = 0$ \\
         5 & $\frac{1}{2}v \Gamma^{\mu}v = 0$ \\
         6 & $\nabla I_3 = 0$ \\
         7 & $\frac{1}{2}\nabla^2 I_4|_{\mathrm{Adj}} = 0$ \\
    \hline
    \end{tabular}
    \caption{The characterizing equations for $v$. For $n=3$, this is the degeneracy condition for a $3\times 2$ matrix. For $n=4$, it is the Plücker relation. For $n=5$, it is the pure spinor condition. For $n=6$, it is the rank-one Jordan algebra orbit condition. For $n=7$, it is the rank-one Freudenthal triple system orbit condition.}
    \label{quadratic-eqs}
\end{table}

As mentioned above, in the cases for $\tilde{T}^n$ with $n=3,4$, we can show that those quadratic equations are exactly equivalent to the level matching condition of orbifolds in a perturbative string frame. Hence, those equations completely characterize anomaly-free orbifolds in those dimensions. In addition, those equations are valid not only for $\mathbb Z_2$ but for general $\mathbb Z_N$. We therefore conjecture \cite{work}: 
\begin{align}
    \boxed{(\text{Orbifold anomaly of } v) \equiv f_i(v) \pmod{N},}
\end{align}
where $ v\in \Lambda_{n,N}^{*}$ is a general $\mathbb Z_N$ phase shift, so that the vanishing of all $f_i$ implies that the orbifold is anomaly-free. Note that this conjecture restricts the class of consistent orbifolds rather than enlarging it. Therefore, the results in this work do not rely on the validity of the conjecture.

\section{Spin moduli space and the spin duality group}\label{section3}
\subsection{Spin duality group}
The notion of generalized spin structure leads to a new moduli space. The original marked moduli space of M-theory compactified on $\tilde{T}^n$ is the homogeneous space
\begin{equation}
    \widehat{\mathcal{M}}_n=\E_{n(n)}(\RR)/\mathrm{H},
\end{equation}
described by the background fields.\footnote{We neglect the $\RR^+$ factor for $n=1,2$ here.} Now there is an additional discrete choice of generalized spin structure, so the corresponding marked space is
\begin{equation}
    \widehat{\mathcal{M}}_n^{\rm spin}=\mathcal{S}_n\times (\E_{n(n)}(\RR)/\mathrm{H}).
\end{equation}
Then we can quotient the marked moduli space by the duality group to get the moduli space of M-theory compactified on $\tilde{T}^{n}$ with a non-trivial generalized spin structure: 
\begin{equation}
    \mathcal{M}_n^{\rm spin}=\E_{n(n)}(\ZZ)\Big\backslash\left(\mathcal{S}_n\times (\E_{n(n)}(\RR)/\mathrm{H})\right).
\end{equation}
We call this the \emph{spin moduli space}. 

Since $\mathcal S_n$ was defined as the U-duality orbit of a geometric shift, the duality group acts transitively on it. Thus, after choosing a specific generalized spin structure $ v\in \mathcal{S}_n$, we then have
\begin{equation}
    \mathcal{S}_n=\frac{\E_{n(n)}(\ZZ)}{\E_{n(n)}^{(v)}(\ZZ)},
\end{equation}
where $\E_{n(n)}^{(v)}$ is the stabilizer of $v$
\begin{equation}
    \E_{n(n)}^{(v)}(\ZZ)=\{ g\in \E_{n(n)}(\ZZ)\mid g\cdot v\equiv v \pmod 2\},
\end{equation}
i.e. the subgroup fixing the spin structure. Therefore, we get
\begin{equation}
    \mathcal{M}_n^{\rm spin}=\E_{n(n)}^{(v)}(\ZZ)\backslash \E_{n(n)}(\RR)/\mathrm{H}.
\end{equation}
Thus, fixing a generalized spin structure enlarges the quotient moduli space by reducing the duality group to the \emph{spin duality group}. Since the U-duality group acts on $\mathcal{S}_n$ transitively, the conjugacy classes of $\E_{n(n)}^{(v)}(\ZZ)$ do not depend on the choice of $v$, so we can denote it by $\E_{n(n)}^{\rm spin}(\ZZ)$. We next give a few examples.
\subsection{9d spin duality group}
For $d=9$ ($n=2$), the U-duality group is just the large diffeomorphism group $\SL(2,\ZZ)$. To get rid of $\RR^+$ one can take the F-theory limit and view this as part of a 0B theory, as was done in \cite{Baykara:2026gem}. Choosing the first circle to have an anti-periodic boundary condition,  the spin-preserving duality group is then
\begin{equation}
\begin{aligned}
    \E_{2(2)}^{\rm spin} &=\left\{ A=\begin{pmatrix}
        a&b\\
        c&d
    \end{pmatrix}\in\SL(2,\ZZ)\Big\vert\, A\begin{pmatrix}
        1\\
        0
    \end{pmatrix}\equiv\begin{pmatrix}
        1\\
        0
    \end{pmatrix}\pmod{2}\right\}\\
    &=\left\{ A=\begin{pmatrix}
        a&b\\
        c&d
    \end{pmatrix}\in\SL(2,\ZZ)\Big\vert\, c\equiv 0\pmod{2}\right\}\\
    &=\Gamma_0(2).
\end{aligned}
\end{equation}
Therefore, the spin moduli space is the moduli curve $Y_0(2)$. 

\subsection{8d spin duality group}
The case of $d=8$ ($n=3$) is more interesting because the T-duality group is non-trivial.

For $n=3$, the U-duality group is $\SL(3,\ZZ)\times \SL(2,\ZZ)$ and the charge lattice transforms in the bi-fundamental representation $(\mathbf 3,\mathbf 2)$ and hence the dual lattice transforms in the dual representation of the form:
\begin{equation}
    \begin{pmatrix}
        u&\omega
    \end{pmatrix}=\begin{pmatrix}
        u^1& \omega_{23}\\
        u^2& \omega_{31}\\
        u^3& \omega_{12}\\
    \end{pmatrix}
\end{equation}
where $u^i$ is the shift dual to the KK momentum in the $i$-th direction and $\omega_{ij}$ is the shift dual to the M2-brane charge wrapping the $ij$ cycles. Choosing an anti-periodic boundary condition for the first cycle, the spin preserving duality group is given by
\begin{equation}
\begin{aligned}
    \E^{\rm spin}_{3(3)}&=\left\{(A,B)\in\SL(3,\ZZ)\times \SL(2,\ZZ)\Bigg\vert\, A\begin{pmatrix}
        1\\0\\0\\
    \end{pmatrix}\equiv \begin{pmatrix}
        1\\0\\0\\
    \end{pmatrix},\, B\begin{pmatrix}
        1\\0
    \end{pmatrix} \equiv \begin{pmatrix}
        1\\0
    \end{pmatrix} \pmod 2\right\}\\
    &=\Gamma^3_0(2)\times \Gamma_0(2),
\end{aligned}
\end{equation}
where we defined
\begin{equation}
    \Gamma_0^3(2)=\left\{ A=\begin{pmatrix}
        a&b&c\\
        d&e&f\\
        g&h&i
    \end{pmatrix}\in \SL(3,\ZZ)\Bigg\vert\, d\equiv g\equiv 0\pmod 2\right\}.
\end{equation}

It is useful to see how this reduction appears from a geometric perspective. The $\SL(3,\ZZ)$ part of the U-duality group is the large diffeomorphism group, so it is clear that the spin preserving symmetry in the $\SL(3,\ZZ)$ should be $\Gamma_0^3(2)$. However, the T-duality $\SL(2,\ZZ)$ part of the duality group is non-geometric in the M-theory picture. A naive geometric guess would be that only the geometric diffeomorphism group acts on the geometric spin structure, and would incorrectly guess that the quantum $\SL(2,\ZZ)$ does not. But from our previous analysis, we know that this $\SL(2,\ZZ)$ should also reduce to $\Gamma_0(2)$. Can we see geometrically how this is possible?

Yes, we just need to go to the dual type IIB picture! First reduce M-theory on the third circle and then T-dualize along the second circle.\footnote{The first circle has an anti-periodic boundary condition, therefore M-theory reduction on it has problems, see \cite{Baykara:2026gem}.} Then the theory will become type IIB on a two torus $T^2_A$ with the anti-periodic boundary condition on the first circle,
\begin{equation}
    \text{M-theory on $S^1_A\times S_{2}^1\times S_{3}^1$}\leftrightarrow \text{type IIA on $S^1_A\times S_2^1$}\leftrightarrow\text{type IIB on $S^1_A\times \tilde S_2^1$},
\end{equation}
where $\tilde S^1_2$ denotes the circle that is T-dualized. 
Then, the quantum $\SL(2,\mathbb{Z})$ symmetry of M-theory becomes the geometric large diffeomorphism group of the two torus in the type IIB frame. As in the case $n=2$, it reduces to $\Gamma_0(2)$. Thus, the generalized spin structure analysis agrees with the geometric description in the duality frame where the relevant symmetry becomes geometric.

\section{Isolated critical points}\label{section4}
Armed with the knowledge of the duality group of Scherk-Schwarz compactifications,  we can try to use this information to constrain the potential $V$ as a function of light moduli fields.

Given a point $x\in \mathcal{M}_n^{\rm spin}$, we can consider the extra gauge symmetry at that point, which is part of the duality group keeping $x$ invariant: 
\begin{equation}
    G_x=\{ g\in \E_{n(n)}^{\rm spin}(\ZZ)|\, g\cdot x=x\}.
\end{equation}
If this group is non-trivial, then $x$ is a \emph{symmetric point} in the moduli space, and $G_x$ will be the \emph{symmetry group} of that point. This means that $G_x$ will be enhanced from a duality group to a gauge group. The existence of the enhanced symmetry will then impose some non-trivial selection rule on $V$.

The relevant special case is when $x$ is an  \emph{isolated symmetric point}.  This means that all linearized scalar moduli near $x$ are charged under $G_x$ so there are no neutral moduli.  Gauge invariance implies that there cannot be any linear term in $V$. Hence:
\\\\
\emph{If $x$ is an isolated symmetric point in the moduli space $\mathcal{M}_n^{\rm spin}$, then $x$ is a critical point for $V$.}
\\\\
Beyond criticality, the same selection rule also constrains the higher-order terms in the potential. Suppose the linearized moduli fields $t^i$ at the symmetric point $x$ transform in a representation
\begin{align}
R=R_1\oplus R_2\oplus \cdots \oplus R_k
\end{align}
of $G_x$, with the $R_i$ irreducible. Then a general $m$-th order term in the scalar potential has the form
\begin{equation}
    V_{i_1\cdots i_m}t^{i_1}\cdots t^{i_m},
\end{equation}
where $V_{i_1\cdots i_m}$ must define an invariant tensor in $\mathrm{Sym}^{m}(R^\vee)$. Hence, the number of independent parameters is the multiplicity of the trivial representation in $\mathrm{Sym}^{m}(R^\vee)$. The decomposition into irreducible representations therefore imposes further constraints on the expansion of $V$.

For example, the absence of the trivial representation among the $R_i$ implies that there is no linear term, which is the criticality condition discussed above. For the quadratic terms, which determine the mass matrix, Schur's lemma implies that cross terms can occur only between isomorphic irreducible representations. In particular, if an irreducible component $R_i$ is not isomorphic to any other $R_j$, then the mass matrix has no cross term between $R_i$ and the other components and acts on $R_i$ by a single scalar. Therefore, all fields in $R_i$ have the same mass eigenvalue, so the whole irreducible component is massive, massless, or tachyonic together.

The rich properties of the symmetry-driven critical points of the potential motivate the search for isolated symmetric points in each dimension. The explicit strategy will be the following:
\begin{itemize}
    \item Since the non-trivial spin structure reduces the duality group, an isolated symmetric point in the spin moduli space $\mathcal{M}_n^{\rm spin}$ must already be an isolated symmetric point in the original moduli space $\mathcal{M}_n$. Therefore, we first choose a representative $x\in \widehat{\mathcal M}_n$ of such a point and find its stabilizer $G'_x$ in the full U-duality group.
    \item We next find subgroups of $G'_x$ for which
    \begin{enumerate}
        \item there is no fixed direction on the tangent space;
        \item there are fixed elements $v$ in $\PP( \Lambda_{n,2}^{*})$;
        \item there exists a fixed $v$ satisfying the quadratic equations; hence $v\in \mathcal{S}_n$.
    \end{enumerate}
    \item If such a subgroup exists, then it will be contained in the symmetry group of the point $(x,v)\in \mathcal{M}_{n}^{\rm spin}$ under which all of the linearized moduli are charged. Thus, $(x,v)$ is an isolated symmetric point in the theory with $\nabla V\big|_{(x,v)}=0$.
    \item Find a U-duality group element mapping $v$ to a geometric spin structure and express the resulting point in terms of the corresponding physical moduli.
\end{itemize}

In the following section, we start by analyzing the simplest isolated symmetric points: the square torus with axions turned off $[\A_1^n]=[\mathrm{Id}]\in \E_{n(n)}/\mathrm{H}$ and get examples in $n=3,5$ and $7$ corresponding to spacetime dimensions $d=8,6$ and $4$. Then, we do a complete search for $d=8,7$ and a partial search for $d=6$.  For $d=8, 6$ we find additional isolated symmetric points, while for $d=7$ we rule out any symmetry-driven critical points for $V$. Finally, we show that, with a different choice of a spin structure, the point $[\A_1^6]\in \E_{6(6)}/\mathrm{H}$ in $d=5$ is also an isolated symmetric point.

\section{Examples}\label{section5}
In this section, we explore examples of critical points in spacetime dimensions from $d=8$ to $d=4$.

In our conventions the moduli denote the values on the covering torus $\tilde{T}^n$, before the quotient that implements the anti-periodic boundary conditions
\begin{align}
    T^n_A = \frac{\tilde{T}^n}{(-1)^{F}\circ T_v }.
\end{align}
For example, a self-dual radius in the covering torus corresponds to half the self-dual radius in the quotient physical compactification torus.

\subsection{The $\mathrm{A}^n_1$ points in even $d$}\label{A^d_1 points}
Consider the point in moduli space where $\tilde{T}^n$ is the orthogonal torus with radii at the self-dual radius from the type IIA frame as reduced on a basic circle. In addition, all axion moduli are set to $0$. We denote this point by the corresponding lattice, $\A^n_1$.

The symmetry group of the $\A_1^n$ point contains a subgroup generated by the geometric symmetry of the orthogonal torus and double T-dualities along a pair of basic circles as
\begin{equation}
    W(E_n) = \SO(n,\ZZ)\bowtie  \ZZ_2^{n-2}
\end{equation}
for $7\geq n\geq 3$. The $n-2$ T-dualities correspond to choosing $2$ basic circles to T-dualize in and $1$ circle to reduce M-theory to type IIA. They are generated by fixing the M-theory circle and the first T-duality circle and choosing the second circle among the remaining $n-2$ circles.

This group is large enough to isolate the identity point: the $\SO(n,\ZZ)$ part fixes all the geometric moduli, and the T-duality part $\ZZ_2^{n-2}$ fixes volume moduli and axions.

Now we consider the action of this group on the generalized spin structures. Consider the frame where the torus is $A_1^n$ and axions are turned off. Suppose we choose a pure geometric shift:
\begin{align}
    v_{\rm geo} &= (u^i=1,\omega_{ij}=0,\sigma_{ijklm}=0,\tau_i=0).
\end{align}
This spin structure corresponds to an anti-periodic boundary condition on half the diagonal circle. It is invariant under the geometric $\SO(n,\ZZ)$ rotations as that would only permute the basic circles under which the diagonal direction is invariant. However, under T-dualities, geometric shift components are exchanged with wrapped M2-brane shift components, and similarly with the higher brane shift components in lower dimensions. Therefore $v_{\rm geo}$ would not be invariant under T-dualities in this frame. 

Let us demonstrate this more explicitly. Suppose we choose a phase shift vector $v_{\rm geo}$ and track it as we apply T-dualities. Reduce on circle $1$ to type IIA. For example, in the $n=5$ case, under T-duality along $2$-$3$ circle directions, KK momenta and string winding charges, which are equivalent to M2-brane charges wrapping $1-2$ and $1-3$ circles exchange. In addition, D0 charge (which is KK momentum in the M-theory circle direction) and D2/M2 wrapping $2-3$ circles also exchange. These imply that the phase shift vector transforms as
\begin{align}
     (\underbrace{1}_{D0},\underbrace{1,1,1,1}_{\mathrm{KK}},0,\dots,0)\mapsto (\underbrace{0}_{D0},\underbrace{0,0,1,1}_{\rm KK},\omega_{12}=1,\omega_{13}=1,\omega_{23}=1,0,\dots,0).
\end{align}
Here, $\omega_{ij}$ denotes the phase shift corresponding to an M2-brane wrapped on the $i$-$j$ two-cycle.

More generally, the Weyl group $W(E_n)$ acts on the entries of $v$ by permutations since each component is a weight vector of the  representation in the $\A^n_1$ case. Therefore, the invariant choice under the whole symmetry group is 
\begin{equation}
    v_{\bf{1}}=(1,1,1,1,1,1,\cdots,1)\in\PP( \Lambda_{n,2}^*).
\end{equation}
After applying the quadratic equations in appendix \ref{Characterizing Equations}, we find that this shift is a generalized spin structure only for odd $n$. 

We now describe the explicit duality frame in which $v$ becomes geometric. If we choose $g\in E_{n(n)}(\ZZ)$ to be the axion shift as
\begin{align}
    g: C_{ijk}\mapsto C_{ijk}+1,
\end{align}
this will map
\begin{equation}
    g \cdot v_{\rm geo}\equiv \begin{cases}v_{\bf{1}}&\text{ if $n$ is odd}\\v_{\rm geo} &\text{ if $n$ is even} \end{cases}\mod 2,
\end{equation}
see appendix C of \cite{Coimbra:2011ky} for transformations of charges under axion shifts. Thus, $v_{\bf{1}}$ can be mapped to the geometric shift $v_{\rm geo}$ if $n$ is odd. So the point $(\A_1^n,v_{\bf{1}})\in\mathcal{M}_{n}^{\rm spin}$ is dual to $(g\cdot \A_{1}^n,g\cdot v_{\bf{1}}=v_{\rm geo})$, where the latter is represented by the moduli fields $(G_{ij}=\delta_{ij},C_{ijk}=1)$.\footnote{The sign of $C_{ijk}$ shift does not matter since the generalized spin structure is defined mod $2$. In particular, we have $g^{-1}\equiv g \pmod{2}$. Therefore $g$ exchanges $v_{\bf 1}$ and $v_{\rm geo}$ in mod $2$ for odd $n$.} This is the same point as before, just a different duality frame.
\\\\
\textsl{The point $(G_{ij}=\delta_{ij},C_{ijk}=1)$ in the covering space with an anti-periodic boundary condition on half of the diagonal circle is an isolated symmetric point for all odd $n$, i.e. for $d=4,6,8$ spacetime dimensions.}
\\\\
In other words, if we compactify M-theory on $\tilde{T}^n$ with $n=3,5,7$ on equal length orthogonal circles, which from type II perspective are at the self-dual radii, with  $C_{ijk}=1$ for all basic 3-cycles of the $\tilde{T}^n$ and then we mod out by a half-shift along the diagonal appended with $(-1)^F$ then  $\nabla V=0$ at this point and it is an isolated critical point.

\subsection{Complete search in 8d}
The moduli space of the $d=8$ ($n=3$) maximal supergravity is
\begin{equation}
\left(\SL(3,\ZZ)\backslash\SL(3,\RR)/\SO(3)\right)\times\left(\SL(2,\ZZ)\backslash\SL(2,\RR)/\SO(2)\right).
\end{equation}
The moduli are given by
\begin{align}
    &\tau = C_{123}+iV,\qquad G\in \SL(3,\RR)/\SO(3)
\end{align}
where $C_{123}$ is the axion and $V$ is the volume of $\tilde{T}^3$, and $G$ is the $\tilde{T}^3$ shape modulus. The field $\tau$ transforms by the M\"obius action of $\SL(2,\ZZ)$, while the shape modulus $G$ transforms under $\SL(3,\ZZ)$.

From the general study of the structure of the space $\SL(n,\ZZ)\backslash \SL(n,\RR)/\SO(n)$ in appendix \ref{appendixB}, we know that each isolated symmetric point can be represented by the shape modulus of an irreducible symmetric lattice in $\RR^n$. Such lattices for $n=2,\,3$ are completely classified:
\begin{itemize}
    \item the isolated symmetric points of $\SL(2,\ZZ)\backslash\SL(2,\RR)/\SO(2)$ are $\A_1^2$ and $\A_2$;
    \item the isolated symmetric points of $\SL(3,\ZZ)\backslash\SL(3,\RR)/\SO(3)$ are $\A_1^3$,\, $\A_3$,\, $\A_3^*$.
\end{itemize}
The Mathematica computation shows that only the symmetric groups of $\A_1^2,\, \A_1^3$ and $\A_3$ have a fixed spin structure. Thus, there are only two isolated symmetric points in the spin moduli space: $(\A_1^3,\A_1^2)$ and $(\A_3,\A_1^2)$.
\subsubsection{ $(\mathrm{A}_1^3,\mathrm{A}_1^2)$ point}
The choice of moduli corresponds to 
\begin{align}
    \tau&=i, \\
    G&=I_3.
\end{align}
The symmetry group fixes the generalized spin structure
\begin{align}
    v_{\bf{1}} = \begin{pmatrix}
        1& 1\\
        1& 1\\
        1& 1
    \end{pmatrix}\in (\mathbf 3,\mathbf 2).
\end{align}
Here, the first column denotes the geometric shift and the second column the phase shift in winding.\footnote{This convention is fixed by requiring the T-duality group $g\in\SL(2,\ZZ)$ to act on the moduli $\tau=C_{123}+iV$ by the Möbius representation and then tracking the momentum and winding elements.} Doing a duality transformation
\begin{align}
    T =\begin{pmatrix}
        1 & 1\\
        0 & 1
    \end{pmatrix}\in \SL(2,\ZZ),
\end{align}
we end up with a geometric spin structure that corresponds to anti-periodic boundary conditions on the diagonal of the basic circles $1,2,3$,
\begin{align}
    v' = v_{\bf{1}}T \equiv \begin{pmatrix}
        1& 0\\
        1& 0\\
        1& 0
    \end{pmatrix} \pmod{2}.
\end{align}
In this duality frame, where the spin structure is purely geometric, the moduli are given as
\begin{align}
    \tau' &= T\cdot \tau =i+1,\\
    G' &= G = I_3.
\end{align}
This is the special case of $n=3$ from our general discussion of the square torus.
\subsubsection{$(\mathrm{A}_3,\mathrm{A}_1^2)$ point}
Now we choose moduli as
\begin{align}
    \tau&=i,\\
    G&=\frac{1}{2^{2/3}}\begin{pmatrix}
        2 & -1 & 0\\
        -1 & 2 & -1\\
        0 & -1 & 2\\
    \end{pmatrix},
\end{align}
which is the Gram matrix of the lattice $\A_3$ normalized to unit volume. The lattice automorphism group is $S_4$, generated by
\begin{equation}
    s=\begin{pmatrix}
        1&-1&0\\
        0&-1&0\\
        0&0&-1
    \end{pmatrix},\,t=\begin{pmatrix}
         0&0&1\\
        -1&0&1\\
        0&-1&1\\
    \end{pmatrix}.
\end{equation}
The fixed generalized spin structure under $S_4\subset \SL(3,\ZZ)$ and $\ZZ_4\subset \SL(2,\ZZ)$ is given as
\begin{align}
    v=\begin{pmatrix}
        1& 1\\
        0& 0\\
        1& 1
    \end{pmatrix}.
\end{align}
Doing a $T\in \SL(2,\ZZ)$ duality transformation, $v'=vT$ becomes a geometric spin structure on the diagonal of $1$ and $3$ basic circles, and $\tau'=i+1$ with $G'=G$ unchanged.
\subsection{Complete search in 7d}
Similarly to the previous case, the moduli space of $d=7$ ($n=4$) is
\begin{equation}
    \SL(5,\ZZ)\backslash\SL(5,\RR)/\SO(5),
\end{equation}
where the isolated symmetric points correspond to irreducible symmetric lattices in $\RR^5$. There are 7 such lattices \cite{ab433c20-1e03-3a2e-b678-9532243847b2} 
\begin{align}
\A_1^5,\quad \A_5,\quad\A_5^*,\quad\D_5,\quad\D_5^*,\quad\A_5^{+2},\quad\A_5^{+3},
\end{align}
where $\A_5^{+2}$ and $\A_5^{+3}$ denote the lattices obtained by gluing $\A_5$ along $2\lambda$ and $3\lambda$ using the discriminant group generator $\lambda\in \A_5^*/\A_5$.

As a result of our complete search among all the subgroups of their symmetry groups, no subgroup satisfies the two conditions simultaneously: either the subgroup is not large enough to isolate the symmetric point, or the subgroup does not fix a generalized spin structure. We provide the search in the supplementary Mathematica notebook. So we get the following conclusion:
\\\\\textsl{There is no isolated symmetric point in $\mathcal{M}_4^{\rm spin}$.}

\subsection{Partial search in 6d}

In $d=6$ ($n=5$), the moduli space of the maximal supergravity is
\begin{equation}
    \SO(5,5,\mathbb Z)\backslash
    \SO(5,5,\mathbb R)/(\SO(5)\times \SO(5)).
\end{equation}
This is also the Narain moduli space of perturbative string theory on $T^5$.
Equivalently, it is the moduli space of positive five-dimensional polarizations of the even unimodular lattice $\mathrm{II}^{5,5}$.
The linearized moduli transform in the representation $({\bf 5},{\bf 5})$
of the compact group $\SO(5)\times \SO(5)$.

The phase shift lattice for generalized Scherk-Schwarz spin structures is the chiral spinor
representation of $\Spin(5,5)$ reduced modulo two,
\begin{align}
    \Lambda^*_{5,2} \simeq {\bf 16}_{\mathbb Z_2}.
\end{align}
In the exterior-algebra model we use
\begin{align}
    S^-=\Lambda^{\rm odd}\mathbb Z_2^5
\end{align}
for the odd chiral spinor. A generalized spin structure is a non-zero element $v\in S^-$
which lies in the U-duality orbit of a geometric half-shift. Equivalently, $v$ is a pure spinor. Thus its Clifford annihilator
\begin{align}
    {\rm Ann}(v)=\{x\in \mathbb Z_2^{10}\mid \Gamma(x)v=0\}
\end{align}
must be a five-dimensional isotropic subspace of the split quadratic space
$\mathbb Z_2^{5,5}$.

We now explain the finite symmetry groups used in the search. Let $L$ be a positive
definite rank-five even lattice, with dual lattice $L^*$ and discriminant group
\begin{align}
    D_L=L^*/L.
\end{align}
The corresponding Narain lattice is obtained by diagonal gluing:
\begin{equation}
    \Gamma_L
    =
    \{(x_L,x_R)\in L^*\oplus L^*
    \mid x_L-x_R\in L\}.
\end{equation}
A pair of automorphisms $(g_L,g_R)\in {\rm Aut}(L)\times {\rm Aut}(L)$ preserves this
diagonal gluing precisely when $g_L$ and $g_R$ induce the same action on $D_L$. Automorphisms of this lattice can only be symmetries of type IIA if
\begin{align}
    \det g_L=\det g_R=1,
\end{align}
since $\det g_L=-1,\det g_R=1$ elements map IIA to IIB, and $\det g_L=\det g_R=-1$ elements can be symmetries of IIA only when combined with an orientifold action, which we do not consider here.

Therefore the point group we use that corresponds to Narain symmetries is
\begin{equation}
    {\rm Point}_+(L)
    =
    \left\{
    (g_L,g_R)\in {\rm Aut}(L)\times {\rm Aut}(L)
    \; \middle| \;
    \det g_L=\det g_R=1,\quad
    g_L|_{D_L}=g_R|_{D_L}
    \right\}.
\end{equation}
Equivalently, it is generated by two types of elements: T-duality type
\begin{equation}
    (k,1),
    \qquad
    k\in {\rm Aut}_D(L)\cap \SO(5),
\end{equation}
and geometric rotation type
\begin{equation}
    (h,h),
    \qquad
    h\in {\rm Aut}(L)\cap \SO(5),
\end{equation}
where ${\rm Aut}_D(L)$ denotes the kernel of the action of ${\rm Aut}(L)$ on $D_L$.
The first type is a purely left-moving asymmetric action, while the second type is the
ordinary diagonal geometric action.

For a fixed pure spinor $v$ and Narain lattice $\Gamma_L$, the symmetry group of the corresponding point in the spin moduli space is
\begin{equation}
    {\rm Stab}(v)
    =
    \{g\in {\rm Point}_+(L)\mid g \cdot v\equiv v \pmod 2\}.
\end{equation}
The criterion for an isolated symmetric point is then
\begin{equation}
    \left({\bf 5},{\bf 5}\right)^{{\rm Stab}(v)}=0.
\end{equation}

The standard examples of Narain lattices are those obtained from the ADE lattices. For rank 5, there are nine such lattices:
\begin{equation}
    \A_1^5,\quad
    \A_1^3\A_2,\quad
    \A_1\A_2^2,\quad
    \A_1^2\A_3,\quad
    \A_2\A_3,\quad
    \A_1\A_4,\quad
    \A_1\D_4,\quad
    \A_5,\quad
    \D_5 .
\end{equation}
For each point, we enumerate the pure-spinor orbits under ${\rm Point}_+(L)$, compute
the stabilizer of each orbit representative, and test whether the stabilizer has any invariant
linearized moduli. We also check explicitly that every successful pure spinor can be mapped
by an integral $SO(5,5,\mathbb Z)$ duality transformation to a pure geometric spin structure.
The explicit stabilizer generators, duality transformations, and the corresponding geometric
frames are given in the supplementary Mathematica notebook.

\begin{table}[t]
\centering
\resizebox{\textwidth}{!}{
\begin{tabular}{c|cccccccccccc}
\hline
\text{Lattice}
& $\A_1^5$
& $\A_1^3\A_2$
& $\A_1\A_2^2$
& $\A_1^2\A_3$
& $\A_2\A_3$
& $\A_1\A_4$
& $\A_1\D_4$
& $\A_5$
& $\D_5$
& $\mathrm{Q}\A_1^3$
& $\mathrm{Q}\A_1\A_2$
& $\mathrm{Q}\A_3$
\\
\hline
$\#v$
& $2$ & $0$ & $3$ & $3$ & $2$ & $0$ & $2$ & $1$ & $2$ & $1$ & $0$ & $1$
\\
\hline
\end{tabular}
}
\caption{
Number of inequivalent generalized spin structures $v$ for which the stabilizer
${\rm Stab}(v)$ fixes no linearized moduli. The notation $\mathrm{Q}\mathfrak{g}$ denotes the Narain lattice obtained from the direct sum of the $\mathbb Z_{12}$ quasicrystal and $\Gamma^{3,3}(\mathfrak{g})$. The count is by pure-spinor orbit representatives under the full symmetry group of the point. Each viable representative gives a distinct isolated symmetric point in the
spin moduli space.
}
\label{tab:d6-spin-results}
\end{table}

An important subtlety is that a single Narain point can admit more than one inequivalent
spin structure leading to an isolated critical point. For example, this occurs for the familiar
$\A_1^5$ point. One viable choice is the previously discussed fully symmetric spinor
\begin{equation}
    v_{\bf 1}
    =
    \sum_i e_i
    +
    \sum_{i<j<k} e_{ijk}
    +
    e_{12345}
    \in \Lambda^{\rm odd}\mathbb Z_2^5 ,
\end{equation}
which is represented in components by $(1,\ldots,1)\in {\bf 16}_{\mathbb Z_2}$.
This is the choice mentioned in Section~\ref{A^d_1 points}.

However, there is also a second inequivalent consistent spin structure at the same
$\A_1^5$ Narain point:
\begin{equation}\label{eq:v-extra-A15}
    v
    =
    e_4+e_5
    +e_{124}+e_{125}+e_{134}+e_{135}+e_{234}+e_{235}.
\end{equation}
Here $e_{ijk}$ denotes the phase-shift component dual to the M2-brane charge wrapping the complementary two-cycle to $(ijk)$.
This spinor is also pure, since using $b=e_4+e_5$ and $\Omega = e_{12}+e_{13}+e_{23}$, we can write it as
\begin{align}
    v=b\wedge (1+\Omega).
\end{align}
The stabilizer has a geometric component consisting of rotations preserving the decomposition $(123)(45)$. For the asymmetric component, define $T_{ij}$ as $(R_{ij},I_5)$, where $R_{ij}$ is a $5\times 5$ diagonal matrix with $-1$ on the $i$ and $j$ entries, and $1$ everywhere else. These correspond to choosing one of the three dualized circles $*(ij)$ to be the M-theory circle, and T-dualizing in the other two.\footnote{Note that it does not matter which one is chosen as the M-theory circle in the remaining $3$ directions. This is a consequence of the geometric and non-geometric subgroups of the 8d U-duality group factorizing as $SL(2,\mathbb Z)\times SL(3,\mathbb Z)$. } Then, the following symmetries fix $v$:
\begin{align}
T_{12}v=T_{13}v=T_{23}v=T_{45}v
    =v.
\end{align}
In particular, they merely permute the terms in \eqref{eq:v-extra-A15}. Combining the geometric symmetries and the T-dualities, all scalars are charged.

This phenomenon occurs more generally in Table~\ref{tab:d6-spin-results}: several ADE Narain points admit multiple inequivalent generalized spin structures, and these should
be regarded as distinct isolated symmetric points. 

The ADE Narain points do not exhaust the isolated symmetric points of the moduli space. There are also quasicrystalline points, where the left
and right projections of $\mathrm{II}^{5,5}$ are not ordinary lattices but quasicrystals
\cite{Harvey:1987da}. These points can have enlarged finite symmetry
groups acting differently on the left and the right.

The quasicrystal relevant for the present $d=6$ search is a $\Gamma^{2,2}$ Narain lattice with a $\mathbb Z_{12}$ quasicrystallographic symmetry. Let
\begin{align}
    \theta &=
    \left(
        R\!\left(\frac{2\pi}{12}\right);
        R\!\left(\frac{10\pi}{12}\right)
    \right)
    \in SO(2)\times SO(2)\subset SO(2,2),\\
    a_1 &= \frac{1}{\sqrt[4]{3}}\left(1,0;1,0\right)\in \mathbb R^{2,2},
\end{align}
then $a_i=\theta^{i-1}a_1$ for $i=1,2,3,4$ generate the quasicrystalline Narain lattice $\Gamma^{2,2}_{\rm Q}$, which can be shown to be even and unimodular.
Note that in the $a_i$ basis, $\theta$ acts integrally. Thus $\theta$ defines the cyclic $\mathbb Z_{12}$ symmetry generating the quasicrystal.

There are other cyclic quasicrystalline blocks \cite{Baykara:2024vss} of signature $(2,4)$ and $(4,4)$ that can also fit in $(5,5)$. However, for these, the associated $\mathbb Z_N$ action always leaves a sublattice $\Gamma^{1,1}\subset \Gamma^{5,5}$
fixed with no room for an additional symmetry to remove the remaining neutral directions. 

It follows that the only cyclic quasicrystal construction that can combine with an ADE factor while leaving no neutral directions is
\begin{equation}
    \Gamma^{5,5}
    =
    \Gamma^{2,2}_{\rm Q}
    \oplus
    \Gamma^{3,3}(\mathfrak g),
\end{equation}
with $\mathfrak g$ of rank $3$ corresponding to
\begin{equation}
    \A_1^3,\qquad \A_1\A_2,\qquad \A_3.
\end{equation}
Once the $\Gamma^{5,5}$ lattice is constructed as above, the computations are similar to the ADE cases.

The results are summarized in Table~\ref{tab:d6-spin-results}. The number displayed is
the number of inequivalent viable pure-spinor orbit representatives for the corresponding Narain point. Thus the search gives $17$ isolated symmetric points in total. Each lattice has a viable choice of $v$ except
\begin{align}
    \Gamma^{5,5}(\A_1^3\A_2),
    \qquad
    \Gamma^{5,5}(\A_1\A_4),\qquad \Gamma^{2,2}_{\rm Q}\oplus \Gamma^{3,3}(\A_1\A_2).
\end{align}
For these points, every pure-spinor stabilizer leaves at least one neutral modulus.

Any additional examples missed by our classification would have to come from non-cyclic quasicrystalline
symmetry groups for which our arguments do not apply, if there are any. It seems plausible that no such exotic points exist, in which case the above classification would be exhaustive for the $d=6$ spin moduli space, though we do not have a proof.

\subsection{Example in 5d}
For $d=5$ ($n=6$), the moduli space of the maximal supergravity is
\begin{align}
    \E_{6(6)}(\mathbb Z)\backslash \E_{6(6)}(\mathbb R) / \USp(8).
\end{align}
The moduli transform as $\mathbf{42}$ of $\USp(8)$, and the charge lattice transforms as the $\mathbf{27}$ of $\E_{6(6)}$.

Because of the computational complexity, doing a scan for $d\leq 5$ can be infeasible. Therefore, in this subsection we discuss a single illustrative example.

The isolated symmetric point example we consider is the identity point $\A_1^{6}$. For this case, as we explained in section \ref{A^d_1 points}, the diagonal choice of $v_{\bf{1}}$ does not work as $n$ is even. However, we will now show that there is a different spin structure choice that satisfies the conditions
\begin{align}
    v = v_{\bf{1}} - (0,\dots,0,\omega_{12}=1,\omega_{34}=1,\omega_{56}=1,0,\dots,0),
\end{align}
with the symmetries given by $W(\mathrm{F}_4)$.

This choice of shift vector can be explained by the structure of the cubic Jordan algebra. The $\mathbf{27}$ representation of the $\E_{6(6)}(\ZZ)$ can be organized into a cubic Jordan algebra $\mathfrak{J}_{\mathbb{O}_s}^3$ (see \cite{Ferrara:1997ci,Ferrara:1997uz,Borsten:2011ai} for the details). There are two invariants associated with the Jordan algebra: a cubic form invariant $I_3$, relevant to the constraining equation of generalized spin structures as explained in appendix \ref{appendixA}, and a choice of an identity element $c\in\mathbf{27}$ of the algebra.

If we require a linear transformation to preserve only the cubic form, the symmetry group is the U-duality group $\E_{6(6)}(\ZZ)$
\begin{equation}
    \E_{6(6)}(\ZZ)=\{g\in\GL(27,\ZZ)|\,I_3(g(\cdot))=I_3(\cdot)\}.
\end{equation}
If we also require the symmetry group to fix the identity element $c$, we then get the exceptional group $\mathrm{F}_{4(4)}$:
\begin{equation}
    \mathrm{F}_{4(4)}(\ZZ)=\{g\in \E_{6(6)}(\ZZ)|\,g\cdot c=c\}.
\end{equation}
This is the standard embedding of $\mathrm{F}_{4(4)}(\ZZ)\hookrightarrow \E_{6(6)}(\ZZ)$. 

The vector 
\begin{align}
    c=(0,\dots,0, \omega_{12}=1,\omega_{34}=1,\omega_{56}=1,0,\dots,0)
\end{align} 
is a valid choice of the identity element. Hence, in particular, the Weyl group $W(\mathrm{F}_{4(4)})$ as a subgroup of $W(\mathrm{E}_{6(6)})$ will fix both $v_{\mathbf{1}}$ and $c$, hence their difference $v=v_{\mathbf{1}}-c$. 

The linearized moduli decompose under the maximal compact subgroup $\USp(6)\times\SU(2)\subset \mathrm{F}_{4(4)}$ as
\begin{equation}
\begin{aligned}
    \USp(8)&\to \USp(6)\times \SU(2)\\
    \mathbf{42} &\to (\mathbf{14},\mathbf{1})\oplus (\mathbf{14'},\mathbf{2}),
\end{aligned}
\end{equation}
with no trivial representation. Equivalently, restricting to the finite symmetry group $W(\mathrm F_4)$ used above, one finds no invariant linearized modulus. Hence the point is an isolated symmetric point.

Now we show that $v$ is a generalized spin structure by showing that it satisfies the quadratic condition \eqref{Jordan-algebra-condition}. For example, the first condition with $a=1,b=2$ gives
\begin{align}
    (\omega_{34}\omega_{56} +\omega_{35}\omega_{46}+\omega_{36}\omega_{45}) +u^1 \sigma^2-u^2\sigma^1 \equiv 0\pmod{2},
\end{align}
and for $a=1,b=3$,
\begin{align}
    (\omega_{24}\omega_{56}+\omega_{25}\omega_{46}+\omega_{26}\omega_{45})+u^1\sigma^3-u^3\sigma^1 \equiv 0\pmod{2},
\end{align}
and all other choices of $a,b$ follow this pattern. For the second condition,
\begin{align}
    \omega_{ai}u^i,
\end{align}
there are a priori $5$ terms, but one of them vanishes since $\omega_{12}=\omega_{34}=\omega_{56}=0$ in our choice. Therefore, the expression evaluates to $4\equiv 0\pmod{2}$. The conditions involving $\sigma^i$ vanish modulo $2$ similarly.

This shows that $v$ is a generalized spin structure. Note that neither $v_{\bf{1}}$ nor $c$ satisfy the condition by themselves, but as we have shown, $v=v_{\bf 1}-c$ does. 

Now we determine values of the moduli fields for the frame in which $v$ is purely geometric. Under axion shifts by $C_3$ and $C_6$, the phase shift $(u,\omega,\sigma)$ transforms as \cite{Coimbra:2011ky}
\begin{align}
    u' &= u,\\
    \omega' &= \omega+\imath_u C_3,\\
    \sigma' &= \sigma+C_3 \wedge \omega +\frac 1 2 C_3 \wedge \imath_u C_3+ \imath_u C_6.
\end{align}
Let
\begin{align}
    v_{\rm geo} = e_1+\cdots +e_6,
\end{align}
with no M2 charge $\omega$ or M5 charge $\sigma$ shifts. Choosing
\begin{align}
    C_3&= e^{135}+e^{146}+e^{236}+e^{245},\\
    C_6 &= 0,
\end{align}
we get the desired phase shift vector
\begin{align}
    v' &\equiv \left(\sum_i e_i\right) + \left(\sum_{i<j} e^{ij}-e^{12}-e^{34}-e^{56}\right)+\sum_i *e^{i}\\
    &\equiv v_{\bf 1} - e^{12}-e^{34}-e^{56}\pmod{2}.
\end{align}
So the frame in which the phase shift is purely geometric is given by moduli
\begin{align}
    G&=I,\\
    C_3 &= e^{135}+e^{146}+e^{236}+e^{245},\\
    C_6 &=0.
\end{align}

\section{Conclusion}\label{section6}
In this paper, we identified the unbroken part of the duality group which survives the Scherk-Schwarz compactification of M-theory on tori and used this to find isolated critical points for the potential as a function of moduli. In particular, we found such points for compactifications to $d=8,\,6,\,5$ and $4$ dimensions.

It would be natural to ask whether one can do a generalized Scherk-Schwarz mechanism on more complicated supersymmetric backgrounds like Calabi-Yau n-folds, for example by orbifolding the Calabi-Yau by a $\ZZ_2$ isometry with no fixed points accompanied by $(-1)^F$.
Indeed, our work motivates the question of whether similar techniques can identify strong-coupling points in moduli space that lead to critical points of the potential.  If we are to start from supersymmetric backgrounds for which we may know the duality group, this would require techniques to find out how the deformations that break the supersymmetry (such as fluxes) affect the duality symmetries.  This is an important direction for future investigation.

We have also found the generalization of level matching for orbifold actions on toroidal compactifications of M-theory realized as phases on the charge lattice.  It would be interesting to extend this to more general U-duality orbifold actions including permutations of charged states as well as to understand the geometric underpinning of this anomaly cancellation condition.

\section*{Acknowledgments}
This work is supported in part by a grant from the Simons Foundation (602883, CV) and the DellaPietra Foundation.

\begin{appendix}
    
\section{Orbit of the geometric spin structure}\label{appendixA}

The goal of this section is to derive the equations characterizing the generalized spin structure in the space $\PP( \Lambda_{n,2}^*)$. As in many cases in mathematics, reduction modulo $2$ introduces a number-theoretic subtlety. This subtlety can be overcome by using algebraic geometry. In comparison to the complicated mathematical details, the conclusion is rather simple and more general:  there is a universal set of equations characterizing the locus of the orbit of the geometric shift vector after reduction modulo $N$.

Therefore, we organize this appendix as follows: in section \ref{Physical Intuition}, we explain the physical intuition behind the set of equations, which have been studied in the context of maximal BPS black holes. Then, in section \ref{Mathematical Preliminary}, we show how to use a technique from algebraic geometry to prove that the same equations can be used in the mod 2 scenario. Readers not interested in the abstract mathematical details may skip ahead to the explicit equations in section \ref{Characterizing Equations}.

In this appendix, we always assume 
\begin{align}
    3\leq n\leq 7.
\end{align}
\subsection{$\frac{1}{2}$-BPS condition}\label{Physical Intuition}
The charge lattice and its dual, which we call the \emph{phase shift lattice}, are representations that are either isomorphic or related by an outer automorphism. In some frame, the highest weight vector corresponds to KK momentum in the charge lattice or the geometric shift in the phase shift lattice. Although there has not been much work on the U-duality orbits of the phase shift lattice, the orbit of the highest weight vector in the charge lattice is well-studied. Thus, we can use the mathematical tools developed in the study of the charge lattice to equivalently study the phase shift lattice.

For M-theory compactified on $T^n$ and a KK-momentum charge vector $q\in \Lambda_{n}$, there is a $\frac{1}{2}$-BPS state with this charge $q$. For example, if the momentum is in the M-theory circle direction, this is the state of coincident D0-branes.

Since the U-duality group maps the $\frac{1}{2}$-BPS states to $\frac{1}{2}$-BPS states, there is a simple necessary condition for a general charge vector $q \in \Lambda_{n}$ to be in the same orbit as the KK momenta:
\\\\
\emph{Given a charge vector $q\in\Lambda_{n}$, if $q$ is in the U-duality orbit of the KK momentum charges, then there must be a $\frac{1}{2}$-BPS state with charge $q$.}
\\\\
In terms of the $d$-dimensional supersymmetry algebra, given a charge vector $q\in \Lambda_n$, there is a moduli-dependent map $Z(\phi,q)$ to the central charge matrices. The existence of $\frac{1}{2}$-BPS states will then impose a constraint on the rank of the central charge matrices which is independent of the moduli. As discussed and proven in \cite{Ferrara:1997ci,Ferrara:1997uz,Borsten:2011ai}, this actually characterizes the orbit of the KK momenta:
\\\\
\emph{A charge vector $q\in\Lambda_n$ is in the U-duality orbit of the KK momenta if and only if there exists a $\frac{1}{2}$-BPS state with charge $q$. Moreover, this property is completely characterized by the rank condition of the central charge matrices.}
\\\\
This gives us the quadratic equations characterizing the orbit of the KK momenta in $\Lambda_n$ for all dimensions, as in Table \ref{quadratic-eqs}. As mentioned at the beginning of this section, the orbit of the geometric shift under the U-duality group is characterized by the equivalent set of equations. The explicit form of the equations will be discussed in subsection \ref{Characterizing Equations}.

In fact, the same set of equations characterizes the orbit after reduction modulo $N$, so, in particular, it characterizes the generalized spin structure. The goal of the next section is to explain why this is true by using the techniques from algebraic geometry.

\subsection{Mod $2$ reduction of U-duality groups}\label{Mathematical Preliminary}
We first give a brief review of the theory of split semi-simple algebraic groups and explain their relation to the U-duality groups. The mathematical details can be found in standard references \cite{SGA3, Milne_2017, Donkin, Platonov_Rapinchuk_Rapinchuk_2023}.

All of the U-duality groups $\E_{n(n)}(\ZZ)$ are Chevalley's integral models of the split semi-simple algebraic groups. This means that most of the information can be extracted by the corresponding discrete root data $\E_{n}$ and there is a universal behavior of the group under the base change:
\begin{equation}
    \E_{n(n)}(\ZZ)\to\E_{n(n)}(R)
\end{equation}
for an arbitrary ring $R$.

In particular, given a dominant integral weight $\lambda$ in the root data, we can define an integral lattice
\begin{equation}
    V_{\ZZ}(\lambda),
\end{equation}
such that for all coefficient rings $R$,
\begin{equation}
    V_{R}(\lambda)=V_{\ZZ}(\lambda)\otimes _{\ZZ}R
\end{equation}
is an $R$ representation of $\E_{n(n)}(R)$. This representation is induced from the maximal torus if we view the weight $\lambda$ as a character, hence a one dimensional representation of the maximal torus of $\E_{n(n)}(R)$. This is sometimes called the standard module with respect to the weight $\lambda$.

If $R$ is a field of characteristic zero, this would be the irreducible representation corresponding to the weight $\lambda$. However, over a field of positive characteristic, which is the mod 2 field $ \ZZ_2$ in our case, the irreducible $R$-representation corresponding to the weight $\lambda$ will in general only be a quotient of $V_R(\lambda)$. Nonetheless, it is proven in \cite{GARIBALDI201769} that all the charge lattices and their dual lattices as representations of the U-duality groups are irreducible after any change of base ring since they are minuscule.\footnote{A representation is minuscule if all of its weights are in one orbit of the Weyl group. This prevents the representation from being further reduced.} Moreover, the geometric shift is a highest weight vector of the representation.

Finally, if we take the simply connected forms of the U-duality groups, the mod $N$ reduction
\begin{equation}
    \E_{n(n)}(\ZZ)\to \E_{n(n)}(\ZZ_N)
\end{equation}
is surjective. So the mod $N$ reduction is the same as changing the base ring.  Hence, the physical question:
\\\\
\emph{Taking the mod $2$ reduction, what is the orbit of the geometric shift $v_{\rm geo}$ in $\PP( \Lambda_{n,2}^*)$ under the action of $\E_{n(n)}(\ZZ)$?}
\\\\
can be translated to the mathematical question
\\\\
\emph{What is the orbit of the highest weight vector $v$ in $\PP(V_{ \ZZ_2}(\lambda))$ under the action of $\E_{n(n)}( \ZZ_2)$?}
\\\\
Here, $\lambda$ is the dominant weight defining the representation. Since the geometric shift $v$ is a highest weight vector, the stabilizer of $v$ in $\PP( V_{\ZZ_2}(\lambda))$
\begin{equation}
    P:=\mathrm{Stab}_{\E_{n(n)}( \ZZ_2)}([v])
\end{equation}
is a parabolic subgroup. Then one can show by algebraic group theory that the embedding of the orbit
\begin{equation}
   f: \E_{n(n)}( \ZZ_2)[v]=G/P\hookrightarrow \PP(V_{ \ZZ_2}(\lambda))
\end{equation}
is a closed immersion. Then by the fundamental theory of projective varieties, the defining ideal, i.e. the equations cutting out the locus, is given by the kernel
\begin{equation}
    I=\bigoplus_{m=0}^{\infty}\ker(\mathrm{Sym}^mV^*_{ \ZZ_2}(\lambda)\to H^0(G/P,f^*\mathcal{O}(m))).
\end{equation}
But by the algebro-geometric version of the Borel–Weil–Bott theorem, we have
\begin{equation}
    H^0(G/P,f^*\mathcal{O}(m))=V_{ \ZZ_2}(m\lambda)^*.
\end{equation}
Therefore the defining ideal is given by
\begin{equation}
    I=\bigoplus_{m=0}^{\infty}\ker(\mathrm{Sym}^mV^*_{ \ZZ_2}(\lambda)\to V_{ \ZZ_2}(m\lambda)^*).
\end{equation}
Equivalently, the orbit is cut out by the condition that, for each $m$, the symmetric power of the vector lies entirely in the highest-weight component of $\mathrm{Sym}^m V$ with highest weight $m\lambda$. Now, we can use the $\ZZ$ model to rewrite this by
\begin{align}
I=\bigoplus_{m=0}^{\infty}\ker(\mathrm{Sym}^mV^*_{\ZZ}(\lambda)\otimes_{\ZZ} \ZZ_2\to V_{\ZZ}(m\lambda)^*\otimes_{\ZZ} \ZZ_2).
\end{align}
Since the representation $V_{\ZZ}(m\lambda)^*$ is a free $\ZZ$ module and the map $\mathrm{Sym}^mV^*_{\ZZ}(\lambda)\to V_{\ZZ}(m\lambda)^*$ is surjective, the Tor group $\mathrm{Tor}_1^{\ZZ}(V_{\ZZ}(m\lambda)^*,\ZZ_2)$ vanishes and hence the tensor product commutes with the kernel:
\begin{equation}
I=
\left(
\bigoplus_{m=0}^{\infty}
\ker\!\left(
\mathrm{Sym}^m V^*_{\ZZ}(\lambda)
\to
V_{\ZZ}(m\lambda)^*
\right)
\right)\otimes_{\ZZ}\ZZ_2
=
I_{\ZZ}\otimes_{\ZZ}\ZZ_2 .
\end{equation}
where
\begin{equation}
    I_{\ZZ}=\bigoplus_{m=0}^{\infty}\ker(\mathrm{Sym}^mV^*_{\ZZ}(\lambda)\to V_{\ZZ}(m\lambda)^*).
\end{equation}
Then, it was proven in \cite{Lakshmibai1998} that this $I_{\ZZ}$ is generated by quadratic elements. Moreover, as a kernel between two free modules, there is no information loss after tensoring with a field, and this will recover the half BPS equations mentioned in the previous section. Hence
\\\\
\emph{By choosing integral generators of the half-BPS quadratic equations, the same set of equations remains valid after reduction modulo $2$.}
\\\\
We need to choose the integral generators since the equations $f_i$ and $2f_i$ define the same vanishing locus over characteristic zero, but $2f_i$ becomes trivial after reduction modulo $2$. Moreover, we do not use any specific properties of $2$ in the discussion of this section; hence, the result is actually true for any mod $N$ reduction.  We conjecture that these equations give the analogue of the anomaly-free condition for $\ZZ_N$ phase-shift orbifolds acting on the M-theory charge lattice on tori \cite{work}.

In the next section, we list a set of integral basis elements of the ideal in each dimension.
\subsection{Quadratic equations for each dimension}\label{Characterizing Equations}
In this section, we list all the equations characterizing the generalized spin structure in the phase shift lattices. All equations are of similar form to the two equations in $d=7$ ($n=4$), and we know that, in that specific dimension, the equations are exactly the anomaly-free conditions. This leads to our conjecture that these equations can be related to some anomaly class in the theory.
\subsubsection{8d: Matrix degeneration}
For $d=8$ ($n=3$), as mentioned above, the shift lattice is dual to the bi-fundamental representation $(\mathbf{3,2})$ of $\SL(3,\ZZ)\times\SL(2,\ZZ)$:
\begin{equation}
    \begin{pmatrix}
        u^1&\omega_{23}\\
        u^2& \omega_{31}\\
        u^3&\omega_{12}
    \end{pmatrix}.
\end{equation}
By elementary linear algebra, this is in the orbit of the geometric shift if and only if this $3\times 2$ matrix has rank $1$, and can be characterized by the quadratic equations:
\begin{equation}
    \omega_{ai}u^i=0,
\end{equation}
for all $a$, which is nothing but requiring all the $2\times 2$ minors to vanish.

A stronger statement in this dimension is that fundamental linear algebra actually tells us every non-zero matrix modulo $2$ is equivalent to one of the following two cases
\begin{equation}
    \begin{pmatrix}
        1&0\\
        0& 0\\
        0&0
    \end{pmatrix},\, \begin{pmatrix}
        1&0\\
        0& 1\\
        0&0
    \end{pmatrix}
\end{equation}
by Gauss elimination. The first one is the geometric boundary condition, and the second one is the anomalous orbifold we mention in the type II perturbative string picture. Hence, the generalized spin structure condition completely characterizes the anomaly-free $\ZZ_2$ shifts in the $\U(1)^{r_3}$ gauge group in this dimension.
\subsubsection{7d: Plücker relations for decomposable forms}
For $d=7$ ($n=4$), the shift lattice is dual to the antisymmetric two-tensor representation $\mathbf{10}$ of $\SL(5,\ZZ)$:
\begin{equation}
    \begin{pmatrix}
        0&u^1&u^2&u^3&u^4\\
        -u^1&0&\omega_{34}&\omega_{42}&\omega_{23}\\
        -u^2&-\omega_{34}&0&\omega_{14}&\omega_{31}\\
        -u^3&-\omega_{42}&-\omega_{14}&0&\omega_{12}\\
        -u^4&-\omega_{23}&-\omega_{31}&-\omega_{12}&0
    \end{pmatrix},
\end{equation}
i.e. it is $Z_{IJ}$ with $I,J=0,\cdots,4$ and $Z_{0i}=u^i$ and $Z_{ij}=\omega_{*ij}$ where $*ij$ is the Hodge dual of the basic two cycle $ij$. From the theory of forms, it can be conjugated to the geometric shift if and only if this two-form is \emph{decomposable} (i.e., it can be written as $v\wedge w$), which can be characterized by the Plücker relations:
\begin{equation}
    Z_{IJ}Z_{KL}-Z_{IK}Z_{JL}+Z_{IL}Z_{JK}=0,
\end{equation}
for all choices of $I,J,K,L$.
We can also write the equations explicitly in terms of the shifts:
\begin{equation}
    \begin{aligned}
        \omega_{ai} u^{i}&=0\\
        \frac{1}{8}\epsilon^{ijkl}\omega_{ij}\omega_{kl}&=0,
    \end{aligned}
\end{equation}
 where the $1/8$ factor just cancels the overcounting of the summation, so the equation is integral. In this case, by the theory of the bilinear form, after the mod 2 reduction, we know all the non-zero two-forms are equivalent to one of the following cases:
\begin{equation}
    \begin{pmatrix}
        0&1&0&0&0\\
        -1&0&0&0&0\\
        0&0&0&0&0\\
        0&0&0&0&0\\
        0&0&0&0&0\\
    \end{pmatrix},\,    \begin{pmatrix}
        0&1&0&0&0\\
        -1&0&0&0&0\\
        0&0&0&1&0\\
        0&0&-1&0&0\\
        0&0&0&0&0\\
    \end{pmatrix}.
\end{equation}
Again, the first one is the geometric shift and the second one is the anomalous orbifold. Therefore, the constraint also completely characterizes the anomaly-free $\ZZ_2$ phase shifts in the $\U(1)^{r_4}$ groups in this dimension.
\subsubsection{6d: Pure spinor}
For $d=6$ ($n=5$), the shift lattice is the spinor representation of $\mathbf{16}$ of $\Spin(5,5,\ZZ)$. In the standard Clifford algebra representation, it can be organized as
\begin{equation}
    v =u^ie_i+\omega_{ij}*(e_i\wedge e_j)+\sigma_{12345}*1,
\end{equation}
with the Clifford algebra
\begin{equation}
    \Gamma^i=e_i\wedge,\qquad \,\Gamma^{i+5}=\iota_{e_i},
\end{equation}
for $i=1,\cdots,5$. The condition for a spinor to be mapped to the geometric shift is the \emph{pure spinor condition}, characterized by
\begin{equation}
    \frac{1}{2}v \Gamma^{\mu}v=0,
\end{equation}
for $\mu=1,\cdots,10$. The product between $v$ and $\Gamma^{\mu}v$ is the standard invariant pairing between the spinor and conjugate spinor, which is nothing but wedging them and taking the top form piece in this representation. The equations can also be written explicitly in terms of the physical shifts as:
\begin{equation}
    \begin{aligned}
        \omega_{ai} u^i&=0\\
        \frac{1}{8}\epsilon^{aijkl}\omega_{ij}\omega_{kl}+\sigma_{12345}u^a&=0,
    \end{aligned}
\end{equation}
for all $a$.
\subsubsection{5d: Cubic Jordan algebra}
For $d=5$ ($n=6$), the phase shift lattice is the $\mathbf{27}$ fundamental representation of $\E_{6(6)}(\mathbb{Z})$.  As explained in \cite{Borsten:2011ai}, this representation can be described using the cubic Jordan algebra $\mathfrak{J}_{\mathbb{O}_s}^3$. Let us make this explicit.

The shift lattice is generated by the dual of $15$ M2-brane charges $\omega_{ij}$ wrapping the basic 2-cycle $ij$ with $i<j$ transform as two-forms, $6$ KK momenta $u_i$ and $6$ M5-brane charges $\sigma^{i}:=\sigma_{*i}$ wrapping the dual cycle of $i$, corresponding to the decomposition of the phase shift lattice under $\SL(6,\ZZ) \subset \E_{6(6)}(\ZZ)$
\begin{align}
    \mathbf{27} \to \mathbf{6} \oplus \mathbf{15} \oplus \mathbf{6}.
\end{align}

There is an invariant integral cubic norm on the lattice
\begin{equation}
    I_3=\frac{1}{24}\epsilon^{ijklmn}\omega_{ij}\omega_{kl}\omega_{mn}+\omega_{ij}u^{i}\sigma^{j}.
\end{equation}
Equivalently, if $u$ is viewed as a vector, $\omega$ as a two-form, and $\sigma$ as the corresponding five-form, this can be written as
\begin{equation}
    I_3=\frac{1}{3!}\omega\wedge\omega\wedge\omega+\iota_u\omega\wedge\sigma.
\end{equation}
With the differential forms expression, this is clearly invariant under the large diffeomorphism group $\SL(6,\ZZ)$. If we choose the first circle as the M-theory circle and perform a double T-duality along circles $2$ and $3$, the net effect on the shift lattice is, for $i,j\neq 1,2,3$:
\begin{equation}
    \begin{aligned}
        u^1&\leftrightarrow \omega_{23}\\
        u^2,u^3&\leftrightarrow \omega_{12},\omega_{13}\\
        \omega_{2i}&\leftrightarrow \omega_{3i}\\
        \omega_{ij}&\leftrightarrow \sigma_{123ij}\\
        \sigma_{12456}&\leftrightarrow \sigma_{13456},
    \end{aligned}
\end{equation}
with $u^i,\omega_{1i},\sigma_{23456}$ invariant. The cubic term is invariant under the action. So, this is indeed an invariant cubic form under the U-duality. In fact, $\E_{6(6)}(\ZZ)$ is precisely the group of integral linear transformations of $\ZZ^{27}$ preserving this cubic form:
\begin{equation}
    \E_{6(6)}(\ZZ)=\{g\in\GL(27,\ZZ)|\, I_{3}(g(\cdot))=I_{3}(\cdot)\}.
\end{equation}

The condition for a phase shift vector $v\in\Lambda_{6,2}^*$ to be in the orbit of the geometric shift is given by
\begin{equation}
    \nabla I_3(v)=0.
\end{equation}
Or explicitly, the equations can be expressed in terms of the physical shifts by:
\begin{equation}\label{Jordan-algebra-condition}
\begin{aligned}
    \frac{1}{8}\epsilon^{abijkl}\omega_{ij}\omega_{kl}+u^a\sigma^b-u^{b}\sigma^{a}&=0\\
    \omega_{ai}u^i&=0\\
    \omega_{ai}\sigma^i&=0,\\
\end{aligned}
\end{equation}
for all $a,b$ not equal to each other.
\subsubsection{4d: Freudenthal triple system}
For $d=4$ ($n=7$), the phase shift lattice is the $\mathbf{56}$ fundamental representation of $\E_{7(7)}(\mathbb{Z})$. As in \cite{Ferrara:1997ci, Ferrara:1997uz, Borsten:2011ai}, this can be organized into the Freudenthal triple system.

The phase shift lattice is generated by the dual of $7$ KK momentum $u^i$, $7$ KK monopole $\tau_i$, $21$ M2-brane charges wrapping the $ij$ cycle $\omega_{ij}$ and $21$ M5-brane wrapping the dual cycle $*(ij)$: $\sigma^{ij}:= \sigma_{*ij}$. We can package these shifts into two antisymmetric tensors with indices $I,J=0,\ldots,7$ as in appendix B of \cite{Kallosh:2012yy}:
\begin{equation}
    X^{0i}=u^i,\, X^{ij}=\sigma^{ij},\,Y_{0i}=\tau_i,Y_{ij}=\omega_{ij}.
\end{equation}

There are two invariant quantities for the Freudenthal triple system. One is the antisymmetric bilinear pairing:
\begin{equation}
    \{v,v'\}=\frac{1}{2}(X^{ij}Y_{ij}'-Y_{ij}X'^{ij}),
\end{equation}
which is the electric-magnetic pairing.
The other one is the invariant quartic form:
\begin{align}\label{freudenthal-quartic}
\begin{split}
    I_4&=X^{IJ}Y_{JK}X^{KL}Y_{KI}-\frac{1}{4}(X^{IJ}Y_{IJ})^2 \\ &+\frac{1}{96}(\epsilon_{IJKLMNOP}X^{IJ}X^{KL}X^{MN}X^{OP}+\epsilon^{IJKLMNOP}Y_{IJ}Y_{KL}Y_{MN}Y_{OP}).
\end{split}
\end{align}
Then the duality group $\E_{7(7)}(\ZZ)$ is exactly the integral linear transformation preserving these two invariants.

The $\frac{1}{2}$-BPS condition is given by:
\begin{equation}
    \nabla^2 I_4|_{Adj}(v)=0,
\end{equation}
which means that the projection of the two-tensor $\nabla^2I_4$ to the adjoint representation of $\E_{7(7)}$ is zero. This can be characterized by the equations:
\begin{equation}
    \begin{aligned}
        X^{IM}Y_{JM}&=0\\
        X^{0M}Y_{0M}=&X^{JM}Y_{JM}\\
        X^{IJ}X^{KL}-X^{IK}X^{JL}+X^{IL}X^{JK}&=\frac{1}{8}\epsilon^{IJKLMNOP}Y_{MN}Y_{OP},
    \end{aligned}
\end{equation}
for all $I,J,K,L$ distinct. We can also write these equations in terms of the physical basis of the shifts:
\begin{equation}
    \begin{aligned}
        \omega_{ai}u^i&=0\\
        \sigma^{ai}\tau_i&=0\\
        \sigma^{ai}\omega_{bi}+u^a\tau_b&=0\\
        u^j\tau_j&=\sigma^{aj}\omega_{aj}\\
        u^{a}\sigma^{bc}+u^{b}\sigma^{ca}+u^c\sigma^{ab}&=\frac{1}{8}\epsilon^{abcijkl}\omega_{ij}\omega_{kl}\\
        \tau_{a}\omega_{bc}+\tau_{b}\omega_{ca}+\tau_c\omega_{ab}&=\frac{1}{8}\epsilon_{abcijkl}\sigma^{ij}\sigma^{kl},
    \end{aligned}
\end{equation}
for all $a,b,c$ distinct.
\section{Isolated symmetric points of $\mathrm{SL}(n,\mathbb Z)\backslash
\mathrm{SL}(n,\mathbb R)/\mathrm{SO}(n)$}\label{appendixB}

The space
\begin{align}
    \mathrm{SL}(n,\mathbb Z)\backslash \mathrm{SL}(n,\mathbb R)/\mathrm{SO}(n)
\end{align}
parametrizes unimodular lattices in $\mathbb R^n$ up to rotation, with the $\SL(n,\ZZ)$ implementing equivalence under lattice basis changes. A point has an enhanced symmetry precisely when the corresponding lattice has a non-trivial rotational symmetry in $\SL(n,\ZZ)$. In other words, these are orbifold points of this space. 

More explicitly, if a point is represented by $X\in \mathrm{SL}(n,\mathbb R)$ and $h\in \mathrm{SL}(n,\mathbb Z)$ fixes it, then
\begin{align}
hX=Xg,\qquad g\in \mathrm{SO}(n),
\end{align}
so that
\begin{align}
g=X^{-1}hX.
\end{align}
Thus the arithmetic symmetry group in $\mathrm{SL}(n,\mathbb Z)$ is conjugate to the ordinary rotational symmetry group of the lattice.

The tangent space at such a point is
\begin{align}
\mathfrak{sl}(n,\mathbb R)/\mathfrak{so}(n),
\end{align}
which may be identified with the space of symmetric traceless tensors on $\mathbb R^n$. If $hX=Xg$, then the linearized moduli transform as
\begin{align}
Y\mapsto \mathrm{Ad}_g Y .
\end{align}
Hence all moduli are charged under the enhanced symmetry group $G$ iff the symmetric traceless representation contains no singlet. This is equivalent to the action of $G$ on $\mathbb R^n$ being irreducible: if the representation is reducible, the invariant orthogonal projection onto a proper invariant subspace gives a nonzero invariant symmetric traceless tensor, while irreducibility rules out such tensors by Schur's lemma.

It is therefore enough to consider maximal irreducible finite subgroups. Indeed, if
$G\subset H\subset \mathrm{SL}(n,\mathbb Z)$ are irreducible, then the unique
$G$-invariant unimodular inner product is also $H$-invariant, so both groups determine
the same lattice point.

Thus isolated symmetric points of
\begin{align}
\mathrm{SL}(n,\mathbb Z)\backslash \mathrm{SL}(n,\mathbb R)/\mathrm{SO}(n)
\end{align}
are obtained from conjugacy classes of maximal finite irreducible subgroups of
$\mathrm{SL}(n,\mathbb Z)$. For each such subgroup, Schur's lemma gives a unique invariant unimodular positive-definite inner product, hence a unique lattice point.

For example, for $n=2$ these are the square and hexagonal lattices, corresponding to $\tau=i$ and $\tau=e^{2\pi i/3}$ points. For $n=3$, the analogous isolated points are the irreducible three-dimensional Bravais lattice points. For the classification of irreducible subgroups in higher dimensions, we refer to \cite{ab433c20-1e03-3a2e-b678-9532243847b2}.

\end{appendix}

\bibliographystyle{JHEP}
\bibliography{refs}
\end{document}